\documentclass[journal]{IEEEtran}
\usepackage[nobreak]{cite}
\usepackage{graphicx}
\usepackage{float}
\usepackage{amsmath}
\usepackage{amssymb}
\usepackage{amsfonts}
\usepackage[printonlyused]{acronym}
\usepackage[capitalise]{cleveref}
\crefformat{equation}{#2(#1)#3}
\Crefformat{equation}{#2(#1)#3}
\usepackage{multirow}

\providecommand{\expe}[1]{\ensuremath{\mathrm{e}^{#1}}}
\def\sinc{\mathop{\mathrm{sinc}}\nolimits}
\providecommand{\abs}[1]{\lvert#1\rvert}

\def\g{\mathop{\mathrm{g}}\nolimits}

\usepackage{url}
\usepackage{tikz}
\usetikzlibrary{calc}

\newcommand{\positiontextbox}[4][]{%
	\begin{tikzpicture}[remember picture,overlay]
		\node[inner sep=3pt, fill=yellow,align=left,draw,line width=1pt,#1] at ($(current page.north west) + (#2,-#3)$) {\parbox{.80\paperwidth}{#4}};
	\end{tikzpicture}%
}

\begin{document}

\title{Time-Modulated Multibeam Phased Arrays with Periodic Nyquist Pulses}	
\author{Roberto Maneiro-Catoira,~\IEEEmembership{Member,~IEEE,}
	Julio Br\'egains,~\IEEEmembership{Senior~Member,~IEEE,}\\
	Jos\'e A. Garc\'ia-Naya,~\IEEEmembership{Member,~IEEE,}
	and~Luis Castedo,~\IEEEmembership{Senior~Member,~IEEE}% <-this % stops a space
	\thanks{%
		$^*$ Corresponding author: Jos\'e A. Garc\'ia-Naya (jagarcia@udc.es).
		
		This work has been funded by the Xunta de Galicia (ED431C 2016-045, ED341D R2016/012, ED431G/01), the Agencia Estatal de Investigaci\'on of Spain (TEC2015-69648-REDC, TEC2016-75067-C4-1-R) and ERDF funds of the EU (AEI/FEDER, UE).
		
		The authors are with the University of A Coru\~na, Spain. E-mail: roberto.maneiro@udc.es, julio.bregains@udc.es, jagarcia@udc.es, luis@udc.es
		
		Digital Object Identifier 10.1109/LAWP.2018.2880087}% <-this % stops a space
	\thanks{}}

\markboth{IEEE ANTENNAS AND WIRELESS PROPAGATION LETTERS}%  
{}
\maketitle

% !TeX spellcheck = en_US
%\begin{acronym}
\acrodef{ADC}[ADC]{analog-to-digital converter}
\acrodef{AWGN}[AWGN]{additive white Gaussian noise}
\acrodef{ASK}[ASK]{amplitude-shift keying}
\acrodef{BER}[BER]{bit error ratio}
\acrodef{BF}[BF]{beamforming}
\acrodef{BFN}[BFN]{beamforming network}
\acrodef{SER}[SER]{symbol error rate}
\acrodef{DAC}[DAC]{digital-to-analog converter}
\acrodef{ETMA}[ETMA]{enhanced time-modulated array}
\acrodef{DC}[DC]{direct current}
\acrodef{DOA}[DOA]{direction of arrival}
\acrodef{DSB}[DSB]{double sideband}
\acrodef{FSK}[FSK]{frequency-shift keying}
\acrodef{FT}[FT]{Fourier Transform}
\acrodef{ISI}[ISI]{inter-symbol interference}
\acrodef{HT}[HT]{Hilbert Transform}
\acrodef{LMD}[LMD]{linearly modulated digital}
\acrodef{LNA}[LNA]{low noise amplifier}
\acrodef{MRC}[MRC]{maximum ratio combining}
\acrodef{MBPA}[MBPA]{Multibeam phased-array antenna}
\acrodef{MSLL}[MSLL]{maximum side-lobe level}
\acrodef{NMLW}[NMLW]{normalized main-lobe width}
\acrodef{NPD}[NPD]{normalized power density}
\acrodef{PCB}[PCB]{printed circuit board}
\acrodef{PS}[PS]{phase shifter}
\acrodef{PSK}[PSK]{phase-shift keying}
\acrodef{QAM}[QAM]{quadrature amplitude modulation}
\acrodef{RF}[RF]{radio frequency}
\acrodef{RFC}[RFC]{Rayleigh fading channel}
\acrodef{RPDC}[RPDC]{reconfigurable power/divider combiner}
\acrodef{SA}[SA]{simulated annealing}
\acrodef{SPMT}[SPMT]{single-pole multiple-throw}
\acrodef{SPST}[SPST]{single-pole single-throw}
\acrodef{SLL}[SLL]{sideband-lobe level}
\acrodef{SR}[SR]{sideband radiation}
\acrodef{SNR}[SNR]{signal-to-noise ratio}
\acrodef{SPDT}[SPDT]{single-pole dual-throw}
\acrodef{SSB}[SSB]{single sideband}
\acrodef{SSB-TM-MBPA}[SSB TM-MBPA]{single sideband time-modulated multibeam phased-array antenna}
\acrodef{SWC}[SWC]{sum of weighted cosines}
\acrodef{STA}[STA]{static array}
\acrodef{TM}[TM]{time-modulated}
\acrodef{TMA}[TMA]{time-modulated array}
\acrodef{TM-MBPA}[TM-MBPA]{time-modulated multibeam phased-array antenna}
\acrodef{VPS}[VPS]{variable phase shifter}
\acrodef{VGA}[VGA]{variable-gain amplifier}
\acrodef{UWB}[UWB]{ultra-wide band}
%\end{acronym}
\begin{abstract}
We present a single sideband time-modulated multibeam phased array governed by periodic Nyquist pulsed signals. A Nyquist pulse is a physically  realizable approach to the ideal sinc function. Hence, its lowpass spectrum suits particularly well for time-modulated arrays to perform harmonic beam steering. Contrarily to switched time-modulated arrays and standard solutions based on variable phase shifters, the performance and complexity of the proposed time modulation scheme is rather robust when increasing the number of multibeams.   
\end{abstract}


\acresetall

% \fi
\IEEEpeerreviewmaketitle
	
	\positiontextbox{11cm}{27cm}{\footnotesize \textcopyright 2018 IEEE. This version of the article has been accepted for publication, after peer review. Personal use of this material is permitted. Permission from IEEE must be obtained for all other uses, in any current or future media, including reprinting/republishing this material for advertising or promotional purposes, creating new collective works, for resale or redistribution to servers or lists, or reuse of any copyrighted component of this work in other works. Published version:
		\url{https://doi.org/10.1109/LAWP.2018.2880087}}
	
\section{Introduction}
\IEEEPARstart{M}{ultibeam} phased-array antennas (\acsp{MBPA}) \acused{MBPA} have multiple steerable directional beams which can either serve a number of distributed users or exploit the multipath channel angular diversity \cite{Hong2017}. Due to its simplicity, an alternative to \acp{MBPA} based on \acp{VPS} is to exploit the different harmonic beam patterns of a switched \ac{TMA} \cite{Li2009PIER,Poli2011,Rocca2014,Maneiro2017_a,ManeiroCatoira2017_Sensors}, which will be referred to as \acused{TM} \ac{TM-MBPA}. However, two key issues inherent to multibeam switched \acp{TMA} clearly compromise the \ac{MBPA} efficiency \cite{Maneiro2018}: 1) the frequency behavior of conventional rectangular pulses, which is not appropriate for efficiently distributing the spectral energy among the multiple harmonic patterns to be exploited; and 2) the duplicated-specular radiation diagrams, which are a consequence of the existence of negative harmonics with the same magnitude and opposite phase. Additionally, the scanning inability of the fundamental mode beam and the proportionality between the phases of harmonics with different order jeopardize the versatility of \acp{MBPA} controlled by conventional \acp{TMA} \cite{Maneiro2018,ManeiroCatoira2018_CommLetters}.

Consequently, a smart use of the beam steering capabilities of switched \acp{TMA}  is relegated to the exploitation of a single harmonic pattern \cite{Amin_Yao2015}. The design of a switched \ac{TM-MBPA} supporting $L$ beams will require $L$ dedicated \acp{BFN}  acting in parallel. As a result, the complexity of the entire \ac{BFN} will increase linearly with $L$ as in standard \acp{MBPA} based on \acp{VPS} \cite{Hong2017}. 

In this letter, we propose an \ac{MBPA} architecture, with a complexity independent of $L$, based on a non-switched \ac{TMA} structure governed by periodic Nyquist pulsed signals, which overcomes the aforementioned \ac{TMA} issues regarding efficiency and flexibility. This is demonstrated by characterizing and quantifying the proposed scheme efficiency. In addition, we consider a \ac{SSB} version of the antenna time modulation operation to remove the duplicated-specular radiation diagrams and further increase the \ac{MBPA} efficiency.
 
\begin{figure}[t]
\centering
\includegraphics[width=\columnwidth]{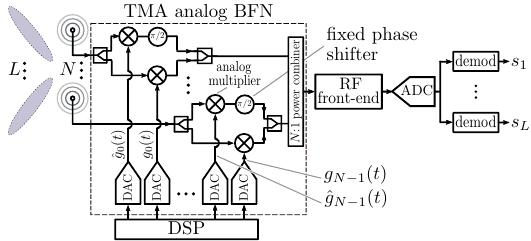}
\caption{Scheme proposed for an \ac{SSB} \ac{TM-MBPA} to handle $L$ beams. The use of a single \ac{BFN} and a single \ac{RF} front-end is a remarkable feature.} 
\label{fig:MBPAs}
\end{figure}

% % % % % % % % % % % % % % % % % % % % % % % % % % % % % % %
\section{Non-Switched SSB TM-MBPA}\label{sec:mathematicalAnalysis}
% % % % % % % % % % % % % % % % % % % % % % % % % % % % % % %
In this section, we introduce the concept of non-switched \ac{SSB-TM-MBPA}, in which time modulation is driven by generic periodic pulsed signals, and derive the time-varying array factors at the different harmonic frequencies.

Let us consider a linear array with $N$ isotropic elements having unitary static excitations $I_n = 1$, $n \in \{0, 1,\dots, N-1\}$. As shown in \cref{fig:MBPAs}, in the \ac{TMA} analog \ac{BFN}, each element excitation is modulated by a pair of periodic pulsed signals: $\g_n(t)$ and its \ac{HT} $\hat{g}_n(t)$, both with  fundamental period $T_0$. We assume that $g_n(t)\in \mathbb{R}$ has no \ac{DC} component and, hence, it can be represented by the following trigonometric Fourier series:
	\begin{equation}\label{eq:Periodic Pulse}
	g_n(t)=2\sum_{q=1}^{\infty}|G_{nq}|\text{cos}(q\omega_o t+\Phi_{nq}),
	\end{equation}
where $\omega_0=2\pi/T_0$ and
	\begin{equation}\label{eq:complex coeff}
	 G_{nq}=|G_{nq}|\expe{j\Phi_{nq}} \in \mathbb{C}
	\end{equation}
	 are the exponential Fourier series coefficients of $g_n(t)$, with modulus $|G_{nq}| \in \mathbb{R}^+$ and phase $\Phi_{nq} \in (0,2\pi]$.
	 Since the $g_n(t)$ are real-valued,  such coefficients satisfy the symmetry property $G_{n(-q)}=G^*_{nq}$. Notice that 
	\begin{equation}\label{eq:HT Periodic Pulse}
	\hat{g}_n(t)=2\sum_{q=1}^{\infty}|G_{nq}|\text{sin}(q\omega_o t+\Phi_{nq}),
	\end{equation}
	because the \ac{HT} produces a $\pi/2$ phase shifting to all the Fourier series components in $g_n(t)$.

The time-varying array factor corresponding to the \ac{SSB-TM-MBPA} with elements treated as shown in \cref{fig:MBPAs} is
\begin{equation}\label{eq:array factor}
F(\theta,t)=\sum_{n=0}^{N-1}[g_n(t)+j\hat{g}_n(t)]\expe{jkz_n\cos\theta},
\end{equation} 
where $z_n$ represents the $n$-th array element position on the $z$ axis, $\theta$ is the angle with respect to such a main axis, and $k$ is the wavenumber. By considering the \ac{FT} of \cref{eq:Periodic Pulse} and \cref{eq:HT Periodic Pulse}, the frequency domain representation of the \ac{SSB-TM-MBPA} \cref{eq:array factor} is
\begin{align}\label{eq:F freq}
F(\theta,\omega)&=\sum_{q=1}^{\infty}\sum_{n=0}^{N-1}2\delta(\omega-q\omega_o)|G_{nq}|\expe{j\Phi_{nq}}\expe{jkz_n\cos\theta},
\end{align}
being $\delta(\omega)$ the unit impulse. Notice that \cref{eq:F freq} does not contain negative spectral lines. 

Turning back to the time domain, we arrive at the following expression for the \ac{SSB-TM-MBPA} time-varying array factor
	\begin{align}\label{eq:F_total}
	F(\theta, t)=\sum_{q=1}^{\infty}F_q(\theta)\expe{jq\omega_0t},
	\end{align} 
where
\begin{equation}\label{eq:F_q}
F_q(\theta)=\sum_{n=0}^{N-1}2G_{nq}\expe{jkz_n\cos\theta}
\end{equation}
are the spatial array factors at frequencies $\omega_c + q \omega_0$,  with $\omega_c$ being the carrier frequency and $q$ positive integer numbers. Notice that such array factors depend on the complex-valued coefficients $G_{nq}$ in \cref{eq:complex coeff} and can be used to design an \ac{MBPA}.

\begin{figure}[t]
	\centering
	\includegraphics[width=\columnwidth]{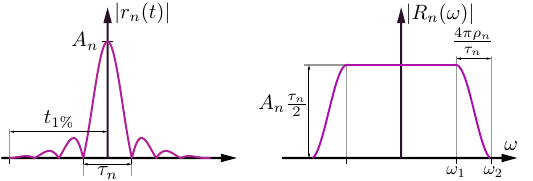}
	\caption{Time \cref{time Nyquist} and frequency \cref{eq:rrcosine} responses of a Nyquist pulse \cite{goldsmith2005wireless}.} 
	\label{fig:Pulso_Base}
\end{figure}
\begin{figure}[h]
	\centering
	\includegraphics[width=\columnwidth]{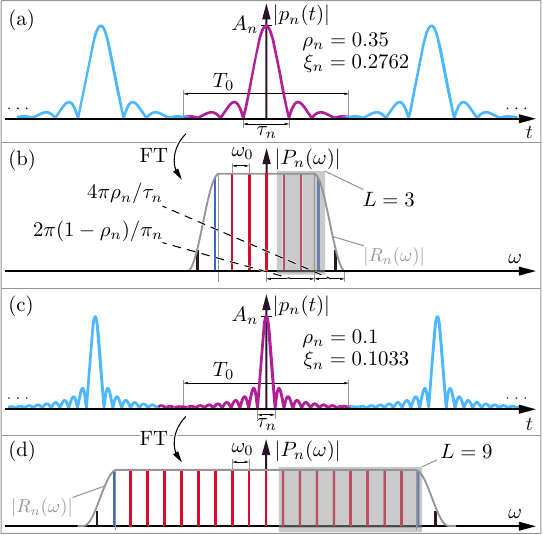}
	\caption{(a) Time response of a periodic Nyquist pulsed signal with $\rho_n=0.35$ and $\xi_n=0.2762$. (b) \ac{FT} of the previous signal which is suitable for windowing $L=3$ harmonic beams (gray-shaded area). (c) Time response of a periodic Nyquist pulsed signal with $\rho_n=0.1$ and $\xi_n=0.1033$. (d) \ac{FT} of the previous signal suitable for windowing $L=9$ harmonic beams.} 
	\label{fig:varios pulsos}
\end{figure}

\section{Control with Periodic Nyquist Pulsed Signals}\label{sec:Nyquist pulses}
The impact of pulse shaping on the sideband radiation of switched \acp{TMA} was studied in \cite{Bekele2013}, whereas the concept of non-switched \ac{TMA} multiple beamforming was introduced in \cite{Maneiro2017_b}. In this section, we particularize the control of the previous generic \ac{SSB-TM-MBPA} to  periodic Nyquist pulses. The long-established alternative to the non-causal ideal $\sinc$ pulse is the well-known Nyquist pulse \cite[Chapter 5]{goldsmith2005wireless}
\begin{equation}\label{time Nyquist}
r_n(t)=A_n\text{sinc}(2\pi t/\tau_n)\frac{\text{cos}(2\rho_n\pi t/\tau_n)}{1- (4\rho_n t/\tau_n)^2},
\end{equation}
where $A_n$ is the amplitude of the pulse at $t=0$, $\tau_n$ is the time separation between zero crossings  and $\rho_n \in [0, 1]$ is the roll-off factor, which determines the smoothness of the pulse frequency response (see \cref{fig:Pulso_Base})
\begin{equation}\label{eq:rrcosine}
%\scriptstyle
\resizebox{0.98\hsize}{!}{$R_n(\omega)= \begin{cases}
A_n{\tau_n}/{2} & \abs{\omega} < \omega_1\\
\displaystyle\frac{A_n\tau_n}{4}\left[ 1 + \cos\displaystyle\left( \frac{\pi(\abs{\omega} - \omega_1)}{{4\pi}/{\tau_n} - 2\omega_1} \right) \right] & \omega_1 \le \abs{\omega} \le \omega_2,
\end{cases}$}
\end{equation}
being $\omega_1 = 2\pi(1-\rho_n)/\tau_n$ and $\omega_2 = 2\pi(1+\rho_n)/\tau_n$.

We next proceed to the construction of the periodic pulsed signals $g_n(t)$ following two steps. We start by constructing $p_n(t)$, the periodic continuation of $r_n(t)$ with fundamental period $T_0$ (see \cref{fig:varios pulsos}), satisfying the constraint
\begin{equation}\label{eq:first restriction}
t_{1\%}\leq T_0/2,
\end{equation}
where $t_{1\%}$ is the time instant beyond which the asymptotic tail of $|r_n(t)|$ is entirely below the 1\% of its maximum level (see \cref{fig:Pulso_Base}). The restriction in \cref{eq:first restriction} is set to guarantee a practically negligible time overlapping between consecutive periods (see Figs.~\ref{fig:varios pulsos}a and \ref{fig:varios pulsos}c), and hence ensuring that the \ac{FT} of $p_n(t)$ is a frequency comb at $\omega=q\omega_0$ ($q \in \mathbb{Z}$), with envelope $R_n(\omega)$ (see Figs.~\ref{fig:varios pulsos}b and \ref{fig:varios pulsos}d).

Thus, the Fourier coefficients of $p_n(t)$ satisfy $P_{nq}=R_n(q\omega_0)/T_0$, i.e.,
\begin{equation}\label{eq:P_nq}
%\scriptstyle
P_{nq}= \begin{cases}
A_n\xi_n/2 &  |q| < \bar{\omega}_1\\
\displaystyle\frac{A_n\xi_n}{4}\left[ 1 + \cos\displaystyle\left( \frac{\pi(|q| - \bar{\omega}_1)}{{2}/{\xi_n} - 2\bar{\omega}_1} \right) \right]& \bar{\omega}_1 \le |q| \le \bar{\omega}_2,
\end{cases}
\end{equation}
with $\bar{\omega}_1=\omega_1/\omega_0$, $\bar{\omega}_2=\omega_2/\omega_0$, and $\xi_n=\tau_n/T_0$. 
We have particularly selected  $t_{1\%}=T_0/2$. Hence, the maximum is $\omega_0$ and the reason is that time modulation is a bandwidth limited technique which must satisfy $\omega_0>B$ \cite{Maneiro2014}, being $B$ the signal bandwidth.

Notice that $T_0$ is usually a fixed parameter in the \ac{TMA} technique. Therefore, for a given $T_0$ (see \cref{fig:varios pulsos}), if $\rho_n$ decreases, $\tau_n$ should also decrease to satisfy the constraint in \cref{eq:first restriction} and, accordingly, $\bar{\omega}_1$ increases. Consequently, the lower $\rho_n$, the higher the number of selected harmonics, and vice versa. More specifically, given an order $L$ of the harmonic beams to be exploited, $\rho_n$ should be selected to satisfy the following requirements: 1) harmonics $|q|\in\{1,\cdots,L-1\}$ are inside the flat zone of the frequency response of $P_n(\omega)$ (see the red spectral lines in Figs.~\ref{fig:varios pulsos}b and \ref{fig:varios pulsos}d) and hence $L-1<\bar{\omega}_1$; 2) harmonics $|q|\in\{L,L+1\}$ are inside the roll-off zone of $P_n(\omega)$, thus the harmonics of order $|q|=L$, the exploited harmonics of highest order (see the blue spectral lines in Figs.~\ref{fig:varios pulsos}b and \ref{fig:varios pulsos}d), suffer a weak attenuation, whereas the ones at $|q|=L+1$ (see the black spectral lines in Figs.~\ref{fig:varios pulsos}b and \ref{fig:varios pulsos}d) experience a significant attenuation to keep the \ac{TMA} efficiency as high as possible. Obviously, the remaining harmonics of higher order are removed. Hence, $\bar{\omega}_1\le L<L+1 \leq \bar{\omega}_2$ and $L+2>\bar{\omega}_2$. 

The second step is to change the phases of the harmonic patterns to endow them with steering capabilities. 
%Recall from \cref{eq:P_nq} that the resulting patterns for $|q|\in\{1,\cdots,L\}$ have uniform distributions. 
This can be done by means of the time delays, $\vartheta_{nq}$, introduced through another periodic signal, $v_n(t)$, also with fundamental period $T_0$. Indeed, let $v_n(t)$ be
\begin{equation}
v_n(t)=\sum_{q=1}^{L_{\text{max}}} \cos\left(\frac{2\pi q}{T_0}\left(t-\vartheta_{nq}\right)\right),
\end{equation}
where $L_{\text{max}}$ is the maximum number of harmonics to be exploited (a fixed bound in the design). The exponential Fourier series coefficient of $v_n(t)$ are $V_{nq}=V_{n(-q)}^*=\expe{-j2\pi q(\vartheta_{nq}/T_0)}$, with $|q|\in [1, L_{\text{max}}]$. If we now construct $g_n(t)$ in \cref{eq:Periodic Pulse} as the periodic convolution
%\begin{equation}\label{eq:periodic conv}
$g_n(t)=v_n(t)\circledast p_n(t)$,
%\end{equation}
the resulting  Fourier series coefficients are
\begin{equation}\label{eq:G_nq final}
G_{nq}=V_{nq}P_{nq} = P_{nq}\expe{-j2\pi q(\vartheta_{nq}/T_0)}.
\end{equation}
Therefore, $\vartheta_{nq}$ determines the phase of the excitation of the $n$-th element of the $q$-th order harmonic pattern.  We remark that $g_n(t)$ and $\hat{g}_n(t)$,  operating in accordance with \cref{fig:MBPAs}, \cref{eq:F_total} and \cref{eq:F_q}, synthesize an \ac{SSB-TM-MBPA}, hence removing the negative harmonic beams. This is illustrated by the gray-shaded areas in Figs.~\ref{fig:varios pulsos}b and \ref{fig:varios pulsos}d with the final useful harmonics appropriately windowed. We also highlight (see \cref{fig:MBPAs}) that the core elements of the scheme are wide-band analog multipliers \cite{Yuwono2009,Makwana2012,Mincica2011,Zhou2005}. The periodic pulses are generated in the digital domain and next converted into the analog domain by means of \acp{DAC}. The \ac{DAC} sampling frequency, $\omega_s$, must be higher than the signal bandwidth \cite{Maneiro2014}, hence becoming independent of $\omega_c$.

\section{Efficiency of the Time Modulation}\label{sec:efficiencies}

In this section, we determine the efficiency of the time modulation operation in the proposed \ac{SSB-TM-MBPA} scheme. 
By virtue of antenna reciprocity, the transmit and receive antenna efficiency is the same. Accordingly, and for the sake of simplicity, we will quantify the efficiency of the time modulation applied to the \ac{SSB-TM-MBPA} in \cref{fig:MBPAs} for the case of transmitting a single carrier with normalized power. Additionally, to simplify the analysis, but without any relevant loss of generality, we will consider a uniform linear array with an inter-element distance of $\lambda/2$. Such an efficiency, denoted by $\eta$, can be factorized in two separated efficiencies
\begin{equation}\label{eq:eficiencias}
\eta=\eta_{\text{TMA}}\cdotp\eta_{\text{mod}}=\frac{P_U^{\text{TM}}}{P_R^{\text{TM}}}\cdotp \frac{P_R^{\text{TM}}}{P_R^{\text{ST}}},
\end{equation}
whose interpretations are described below.

The efficiency $\eta_{\text{TMA}}$ accounts for the competence of the \ac{TMA} technique to filter out and radiate only the working (or useful) harmonics. It is determined by $\eta_{\text{TMA}}=P_U^{\text{TM}}/P_R^{\text{TM}}$, where $P_R^{\text{TM}}$ is the total mean radiated power by the \ac{SSB-TM-MBPA}, which is given by the expression \cite[(16)]{Maneiro2017_a} $P_R^{\text{TM}}=\sum_{q=1}^{L+1}p_q$, being $p_q$ the total mean radiated power at the $q$-th harmonic. On the other hand, $P_U^{\text{TM}}$ denotes the useful mean radiated power (since only $L$ harmonics are profitably exploited) and obeys to  $P_U^{\text{TM}}=\sum_{q=1}^{L}p_q$ \cite[Eq. 16]{Maneiro2017_a}. It is remarkable that the \ac{SSB} operation duplicates the value of this efficiency with respect to that of a conventional \ac{TMA}. 

Most of the works in the literature which analyze the \ac{TMA} efficiency limit themselves to the study of $\eta_{\text{TMA}}$. However, the second component of the efficiency, $\eta_{\text{mod}}$, accounts for the reduction of the total mean power radiated by a uniform static array caused by the effect of modulating its excitations with periodic pulses. $\eta_{\text{mod}}$ is of critical importance due to its high impact on the antenna gain. This efficiency is evaluated by means of the quotient $\eta_{\text{mod}}=P_R^{\text{TM}}/P_R^{\text{ST}}$, where $P_R^{\text{ST}}$ is the total mean power radiated by a uniform static array with $N$ elements and is calculated as the total mean transmitted power over the array factor $F^{\text{ST}}(\theta)=\sum_{n=0}^{N-1}\expe{jkz_n\cos\theta}$, i.e., $P_{\text{R}}^{\text{ST}}= \int_{0}^{2\pi}\int_{0}^{\pi} |F^{\text{ST}}(\theta)|^2\sin(\theta)d\theta d\varphi=4\pi N. $

In order to derive closed-form expressions for both efficiencies, we still have to obtain the expressions of $P_R^{\text{TM}}$ and $P_U^{\text{TM}}$. In this sense, since $p_q$ is given by $p_q=4\pi \sum_{n=0}^{N-1}|G_{nq}|^2$ \cite{Maneiro2017_a}, bearing in mind \cref{eq:P_nq} and \cref{eq:G_nq final}, and selecting $\rho_n$ as specified in \cref{sec:Nyquist pulses}, we have 
\begin{align*}
P_U^{\text{TM}} &= 4\pi N\Bigg[(L-1)\Big(\frac{A_n\xi_n}{2}\Big)^2+\\
&+\Bigg[\frac{A_n\xi_n}{4}\Bigg(1+ \text{cos}\Bigg(\frac{\pi(L-\bar{\omega}_1)}{\frac{2}{\xi_n}-2\bar{\omega}_1}\Bigg)\Bigg)\Bigg]^2\Bigg],\\
P_R ^{\text{TM}}&=P_U^{\text{TM}}+p_{L+1}=P_U^{\text{TM}}+\\
&+4\pi N\Bigg[\frac{A_n\xi_n}{4}\Bigg(1+\text{cos}\Bigg(\frac{\pi((L+1)-\bar{\omega}_1)}{\frac{2}{\xi_n}-2\bar{\omega}_1}\Bigg)\Bigg)\Bigg]^2.
\end{align*}
%and 
%\begin{align}\label{eq:P_R}
%
%\end{align}
Finally, by substituting $P_{\text{R}}^{\text{ST}}$, $P_U^{\text{TM}}$ and $P_R ^{\text{TM}}$ into \cref{eq:eficiencias}, we obtain the aforementioned closed-form expressions.

\begin{table}[t]
\caption{Parameters of Nyquist pulses in \cref{fig:Pulso_Base,fig:varios pulsos}, peak levels of the harmonics inside the roll-off zone, and time modulation efficiency ($\eta_{\mathrm{mod}} \approx 100\%$, hence $\eta = \eta_{\mathrm{TMA}}$ in this case).}
\label{tab:Parameters simulations}
\begin{center}
	\setlength\tabcolsep{0.2em}
	\def\arraystretch{1.1}
	\begin{tabular}{|c|c|c|c|c|c|c|c|c|c|}
		\hline
		{$L$}&{$A_n$\,[dBV]}&{$\rho_n$}&{$\xi_n$}&{${\rho_n}/{\xi_n}$}&{$\bar{\omega}_1$}&{$\bar{\omega}_2$}&{ $L$\,[dB]}&{$L$+1\,[dB]}&{$\eta$ [\%]}\\
		\hline \hline
		3&4.52&0.35&0.28&1.27&2.35&4.89&-1.44&-11.28&98.86\\ \hline
		5&4.99&0.21&0.18&1.18&4.42&6.78&-1.32&-12.20&99.52\\ \hline
		9&5.49&0.10&0.10&0.97&8.73&10.67&-0.42&-11.49&99.76\\ \hline
		15&6.22&0.05&0.06&0.79&14.92&16.49&-0.06&-13.25&99.92\\ \hline
	\end{tabular}
\end{center}
\end{table}

\begin{table}[t]
\caption{Comparison of multibeam  switched \acp{TMA} in the literature to the proposed \ac{SSB-TM-MBPA} ($N$=20 in all cases, $D$ is the mean directivity of the beams).}
\label{tab:Figures of merit}
\begin{center}
	\setlength\tabcolsep{0.2em}
	\def\arraystretch{1.1}
	\begin{tabular}{|c|c|c|c|c|c|}
		\hline
		\multirow{2}{*}{reference} & number & \multirow{2}{*}{$D$\,[dBi]} & \multirow{2}{*}{$\eta_{\text{TMA}}$\,[\%]} & \multirow{2}{*}{$\eta_{\text{mod}}$\,[\%]} & \multirow{2}{*}{$\eta$\,[\%]}\\
		& of BFNs  & & & & \\
		\hline \hline
		\cite{Rocca2014} 	 & 1 &  2.64 & 30.78 &  9.96 &  3.06\\
		\hline
		\cite{Maneiro2017_a} & 1 &  4.79 & 54.75 & 23.96 & 13.12\\
		\hline
		\cite{Amin_Yao2015}  & 3 & 8.22 & 99.79 & 55.55 & 55.44\\
		\hline
		proposed SSB  & \multirow{2}{*}{1} & \multirow{2}{*}{4.37} & \multirow{2}{*}{98.86} & \multirow{2}{*}{100} & \multirow{2}{*}{98.86}\\ 
		TM-MBPA & & & & & \\
		\hline
	\end{tabular}
\end{center}
\end{table}

\section{Numerical Examples}

\cref{tab:Parameters simulations} details, for $T_0=2t_{1\%}$, several configuration examples of an \ac{SSB-TM-MBPA} with $N$=20 elements time-modulated by identical pulses (same $\rho_n$ and $A_n$). The selected values of $\rho_n$ (leading to the corresponding values of $\xi_n$) allow for exploiting $L$= 3, 5, 9 and 15 harmonics, respectively. According to \cref{sec:Nyquist pulses}, such values of $\rho_n$ locate, in each case, $L-1$ harmonics in the flat zone of the frequency response of the pulses, while the harmonics $L$ and $L+1$ are located at the beginning and at the end of the roll-off zone (see $\bar{\omega}_1$, $\bar{\omega}_2$, the attenuation of such harmonics in \cref{tab:Parameters simulations}, and \cref{fig:varios pulsos}). Notice that the parameter $\rho_n/\xi_n$ is directly proportional to the roll-off zone width. Hence, \cref{tab:Parameters simulations} shows that when $L$ increases, the width of roll-off zone decreases. Consequently, since the separation in frequency between adjacent harmonics is fixed ($\omega_0$), when $L$ increases, harmonics $L$ and $L+1$ are better adjusted to the borders of the roll-off zone, and hence $\eta_{\text{TMA}}$ improves. However, if $L$ increases, $\xi_n$ decreases and, if $A_n$ remains fixed, $\eta_{\text{mod}}$ also decreases and consequently $\eta$ also does. Although $\eta_{\text{TMA}}$ is independent of $A_n$ for identical pulses, $\eta_{\text{mod}}$ can be improved by means of adjusting $A_n$, which is not possible in switched \acp{TMA}. \cref{tab:Parameters simulations}  shows the different values of $A_n$ yielding $\eta_{\text{mod}} \approx 100\%$ in each case, and therefore $\eta=\eta_{\text{TMA}}$. 

For the case $L$=3 (there are no examples in the literature with $L>$3), it is possible to perform a comparison with switched \acp{TMA}. Multibeam \acp{TMA} with rectangular pulses implemented with \ac{SPST} switches are bound to use $\xi_n$ very close to zero. Additionally, such \acp{TMA} have no \ac{SSB} features (see \cref{tab:Figures of merit}, \cite{Rocca2014}), leading to modest values for $\eta_{\mathrm{TMA}}$ and $\eta_{\mathrm{mod}}$ and, consequently, low $\eta$ and $D$ (mean directivity of the beams). For \ac{SPDT} schemes, the figures of merit are improved but the frequency response of rectangular pulses together with the lack of \ac{SSB} features still constitute a serious bottleneck (\cref{tab:Figures of merit}, \cite{Maneiro2017_a}). One way of improving the performance is to focus on the generation of a single beam  by using trapezoidal pulses and endowing the \ac{TMA} with \ac{SSB} features \cite{Amin_Yao2015}. Consequently, it is necessary to consider $L$ independent time-modulated \acp{BFN}. However, as the pulses remain at zero state for a certain amount of time, $\eta_{\mathrm{mod}}$ cannot achieve the full level of $\eta_{\mathrm{TMA}}$  (\cref{tab:Figures of merit}, \cite{Amin_Yao2015}). A way of improving  $\eta_{\mathrm{mod}}$ is by amplifying the static excitations (amplifiers at $\omega_c$). Consequently, two key benefits of the proposed non-switched \ac{TMA} scheme are: (1) the safeguard of $\eta_{\mathrm{mod}}$ by amplifying at $\omega_0\ll \omega_c$; and (2) the use of a single \ac{BFN}.

\begin{figure}[t]
	\centering
	\includegraphics[width=\columnwidth]{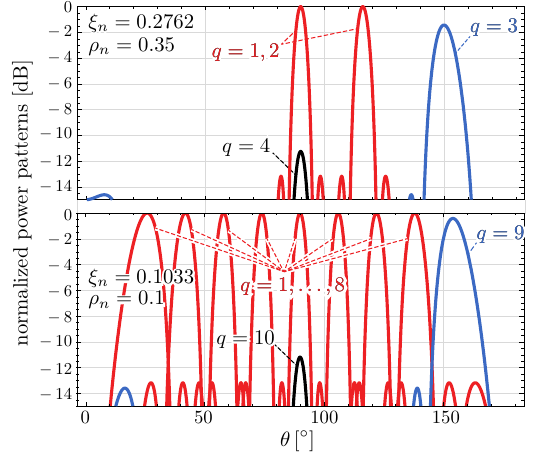}
	\caption{Normalized power radiated patterns of the proposed \ac{SSB-TM-MBPA} with $N$=20 elements after applying the periodic Nyquist pulsed signals shown in \cref{fig:varios pulsos}a (top) and \cref{fig:varios pulsos}c (bottom). The relative levels of the harmonics inside the roll-off zone (blue and black) are specified in \cref{tab:Parameters simulations}. Notice that the phase excitations are considered to be progressive. Hence (see \cref{eq:G_nq final}), a given harmonic beam $q \in \{1, 2,\dots, L\}$ will point to the direction $\theta_q$ when $\vartheta_{nq}/T_0=n\cos(\theta_q)/q$, $n \in \{0, 1,\dots, N-1\}$, whereas $\theta_{L+1}=90^{\circ}$.} 
	\label{fig:radiated patterns}
\end{figure}

\cref{fig:radiated patterns} shows the radiated patterns for the proposed \ac{SSB-TM-MBPA} and the cases $L$=3 (top) and $L$=9 (bottom), respectively. Notice the effect of narrowing the roll-off zone for $L$=9, yielding a better filtering of the exploited higher order harmonics. 

\section{Conclusions}
We have presented a novel \ac{MBPA} scheme based on time modulation with periodic Nyquist pulsed signals and \ac{SSB} operation. The proposed \ac{SSB-TM-MBPA} provides excellent levels of power efficiency and reconfigurability. Furthermore, both the performance  and the \ac{MBPA} \ac{BFN} complexity are invariant to the number of multibeams. 
\ifCLASSOPTIONcaptionsoff
  \newpage
\fi

\vfill



\bibliographystyle{IEEEtran}
\bibliography{IEEEabrv,main}



\vfill


\end{document}